\documentclass[twocolumn,aps,superscriptaddress,showpacs]{revtex4}
\usepackage{amsmath,bm}
\usepackage{graphicx}

\begin{document}

\title{Effects of isospin and momentum dependent interactions on liquid-gas
phase transition in hot asymmetric nuclear matter}
\author{Jun Xu}
\affiliation{Institute of Theoretical Physics, Shanghai Jiao Tong University, Shanghai
200240, China}
\author{Lie-Wen Chen}
\affiliation{Institute of Theoretical Physics, Shanghai Jiao Tong University, Shanghai
200240, China}
\affiliation{Center of Theoretical Nuclear Physics, National Laboratory of Heavy-Ion
Accelerator, Lanzhou, 730000, China}
\author{Bao-An Li}
\affiliation{Department of Physics, Texas A\&M University-Commerce, Commerce, TX
75429-3011, USA}
\author{Hong-Ru Ma}
\affiliation{Institute of Theoretical Physics, Shanghai Jiao Tong University, Shanghai
200240, China}

\begin{abstract}
The liquid-gas phase transition in hot neutron-rich nuclear matter is
investigated within a self-consistent thermal model using an isospin and
momentum dependent interaction (MDI) constrained by the isospin diffusion
data in heavy-ion collisions, a momentum-independent interaction (MID), and
an isoscalar momentum-dependent interaction (eMDYI). The boundary of the
phase-coexistence region is shown to be sensitive to the density dependence
of the nuclear symmetry energy with a softer symmetry energy giving a higher
critical pressure and a larger area of phase-coexistence region. Compared
with the momentum-independent MID interaction, the isospin and
momentum-dependent MDI interaction is found to increase the critical
pressure and enlarge the area of phase-coexistence region. For the isoscalar
momentum-dependent eMDYI interaction, a limiting pressure above which the
liquid-gas phase transition cannot take place has been found and it is shown
to be sensitive to the stiffness of the symmetry energy.
\end{abstract}

\pacs{21.65.+f, 21.30.Fe, 24.10.Pa, 64.10.+h}
\maketitle

The van der Waals behavior of the nucleon-nucleon interaction is expected to
lead to the so-called liquid-gas (LG) phase transition in nuclear matter.
Since the early work, see, e.g., Refs. \cite{lamb78,fin82,ber83,jaqaman83},
many investigations have been carried out to explore properties of the
nuclear LG phase transition both experimentally and theoretically over the
last three decades. For a recent review, see, e.g., Refs. \cite%
{chomaz,das,wci}. Most of these studies focused on investigating features of
the LG phase transition in symmetric nuclear matter. Recent progress in
experiments with radioactive beams provided us a great opportunity to
explore more extensively the LG phase transition in isospin-asymmetric
nuclear matter. New features of the LG phase transition in asymmetric
nuclear matter are expected. In particular, in an asymmetric nuclear matter,
there are two components of protons and neutrons, and two conserved charges
of baryon number and the third component of isospin, and the LG phase
transition was suggested to be of second order \cite{muller95}. This
suggestion together with the need to understand better properties of
asymmetric nuclear matter relevant for both nuclear physics and astrophysics
have stimulated a lot of work recently, see, e.g., Refs. \cite%
{liko97,wang00,su00,lee01,li01,natowitz02,li02,chomaz03,sil04,chomaz06}.
While significant progress has been made recently, many interesting
questions about the nature of the LG phase transition in asymmetric nuclear
matter remain open. Some of these questions can be traced back to our poor
understanding about the isovector nuclear interaction and the density
dependence of the nuclear symmetry energy \cite{wci,ireview98,ibook}.
Fortunately, recent analyses of the isospin diffusion data in heavy-ion
reactions have allowed us to put a stringent constraint on the symmetry
energy of neutron-rich matter at sub-normal densities \cite%
{betty04,chen05,lichen05}. It is therefore interesting to investigate how
the constrained symmetry energy may allow us to better understand the LG
phase transition in asymmetric nuclear matter. Moreover, both the isovector
(i.e., the nuclear symmetry potential) and isoscalar parts of the single
nucleon potential should be momentum dependent due to the non-locality of
nucleon-nucleon interaction and the Pauli exchange effects in many-fermion
systems. However, effects of the momentum-dependent interactions on the LG
phase transition in asymmetric nuclear matter have received so far little
theoretical attention. A timely remedy of the situation is imperative given
the experimental studies underway or planned at various radioactive beam
facilities.

In the present work, we study effects of isospin and momentum dependent
interactions on the LG phase transition in hot neutron-rich nuclear matter
within a self-consistent thermal model using three different interactions.
The first one is the isospin and momentum dependent MDI interaction
constrained by the isospin diffusion data in heavy-ion collisions. The
second one is a momentum-independent interaction (MID) which leads to a
fully momentum independent single nucleon potential, and the third one is an
isoscalar momentum-dependent interaction (eMDYI) in which the isoscalar part
of the single nucleon potential is momentum dependent but the isovector part
of the single nucleon potential is momentum independent.

In the isospin and momentum-dependent MDI interaction, the potential energy
density $V_{\text{MDI}}(\rho ,T,\delta )$ of a thermally equilibrated
asymmetric nuclear matter at total density $\rho $, temperature $T$ and
isospin asymmetry $\delta $ is expressed as follows~\cite{das03,chen05},
\begin{eqnarray}
V_{\text{MDI}}(\rho ,T,\delta ) &=&\frac{A_{u}\rho _{n}\rho _{p}}{\rho _{0}}+%
\frac{A_{l}}{2\rho _{0}}(\rho _{n}^{2}+\rho _{p}^{2})+\frac{B}{\sigma +1}%
\frac{\rho ^{\sigma +1}}{\rho _{0}^{\sigma }}  \notag \\
&\times &(1-x\delta ^{2})+\frac{1}{\rho _{0}}\sum_{\tau ,\tau ^{\prime
}}C_{\tau ,\tau ^{\prime }}  \notag \\
&\times &\int \int d^{3}pd^{3}p^{\prime }\frac{f_{\tau }(\vec{r},\vec{p}%
)f_{\tau ^{\prime }}(\vec{r},\vec{p}^{\prime })}{1+(\vec{p}-\vec{p}^{\prime
})^{2}/\Lambda ^{2}}.  \label{MDIV}
\end{eqnarray}%
In the mean field approximation, Eq. (\ref{MDIV}) leads to the following
single particle potential for a nucleon with momentum $\vec{p}$ and isospin $%
\tau $ in the thermally equilibrated asymmetric nuclear matter \cite%
{das03,chen05}

\begin{eqnarray}
U_{\text{MDI}}(\rho ,T,\delta ,\vec{p},\tau ) &=&A_{u}(x)\frac{\rho _{-\tau }%
}{\rho _{0}}+A_{l}(x)\frac{\rho _{\tau }}{\rho _{0}}+B(\frac{\rho }{\rho _{0}%
})^{\sigma }  \notag \\
&\times &(1-x\delta ^{2})-8\tau x\frac{B}{\sigma +1}\frac{\rho ^{\sigma -1}}{%
\rho _{0}^{\sigma }}\delta \rho _{-\tau }  \notag \\
&+&\frac{2C_{\tau ,\tau }}{\rho _{0}}\int d^{3}p^{\prime }\frac{f_{\tau }(%
\vec{r},\vec{p}^{\prime })}{1+(\vec{p}-\vec{p}^{\prime })^{2}/\Lambda ^{2}}
\notag \\
&+&\frac{2C_{\tau ,-\tau }}{\rho _{0}}\int d^{3}p^{\prime }\frac{f_{-\tau }(%
\vec{r},\vec{p}^{\prime })}{1+(\vec{p}-\vec{p}^{\prime })^{2}/\Lambda ^{2}}.
\label{MDIU}
\end{eqnarray}%
In the above $\tau =1/2$ ($-1/2$) for neutrons (protons) and $f_{\tau }(\vec{%
r},\vec{p})$ is the phase space distribution function at coordinate $\vec{r}$
and momentum $\vec{p}$. The detailed values of the parameters $\sigma
,A_{u}(x),A_{l}(x),B,C_{\tau ,\tau },C_{\tau ,-\tau }$ and $\Lambda $ can be
found in Refs. \cite{das03,chen05} and have been assumed to be temperature
independent here. The isospin and momentum-dependent MDI interaction gives
the binding energy per nucleon of $-16$ MeV, incompressibility $K_{0}=211$
MeV and the symmetry energy of $31.6$ MeV for cold symmetric nuclear matter
at saturation density $\rho _{0}=0.16$ fm$^{-3}$ \cite{das03}. The different
$x$ values in the MDI interaction are introduced to vary the density
dependence of the nuclear symmetry energy while keeping other properties of
the nuclear equation of state fixed \cite{chen05}. We note that the MDI
interaction has been extensively used in the transport model for studying
isospin effects in intermediate energy heavy-ion collisions induced by
neutron-rich nuclei \cite%
{li04b,chen04,chen05,lichen05,li05pion,li06,yong061,yong062}. In particular,
the isospin diffusion data from NSCL/MSU have constrained the value of $x$
to be between $0$ and $-1$ for nuclear matter densities less than about $%
1.2\rho _{0}$ \cite{chen05,lichen05}, we will thus in the present work
consider the two values of $x=0$ and $x=-1$. We note that the
zero-temperature symmetry energy for the MDI interaction with $x=0$ and $-1$
can be parameterized, respectively, as $31.6(\rho /\rho _{0})^{0.69}$ MeV
and $31.6(\rho /\rho _{0})^{1.05}$ MeV \cite{chen05}, and thus $x=0$ gives a
softer symmetry energy while $x=-1$ gives a stiffer symmetry energy.

In the momentum-independent MID interaction, the potential energy density $%
V_{\text{MID}}(\rho ,\delta )$ of a thermally equilibrated asymmetric
nuclear matter at total density $\rho $ and isospin asymmetry $\delta $ can
be written as
\begin{equation}
V_{\text{MID}}(\rho ,\delta )=\frac{\alpha }{2}\frac{\rho ^{2}}{\rho _{0}}+%
\frac{\beta }{1+\gamma }\frac{\rho ^{1+\gamma }}{{\rho _{0}}^{\gamma }}+{%
\rho }E_{sym}^{pot}(\rho ,x){\delta }^{2}.
\end{equation}%
The parameters $\alpha $, $\beta $ and $\gamma $ are determined by the
incompressibility $K_{0}$ of cold symmetric nuclear matter at saturation
density $\rho _{0}$ \cite{liko97}
\begin{eqnarray}
\alpha &=&-29.81-46.90\frac{K_{0}+44.73}{K_{0}-166.32}~\text{(MeV)} \\
\beta &=&23.45\frac{K_{0}+255.78}{K_{0}-166.32}~\text{(MeV)} \\
\gamma &=&\frac{K_{0}+44.73}{211.05}
\end{eqnarray}%
and $K_{0}$ is again set to be $211$ MeV as in the MDI interaction. To fit
the MDI interaction at zero temperature, the potential part of the symmetry
energy $E_{sym}^{pot}(\rho ,x)$ is parameterized by \cite{chen05}
\begin{equation}
E_{sym}^{pot}(\rho ,x)=F(x)\frac{\rho }{\rho _{0}}+\left[ 18.6-F(x)\right] (%
\frac{\rho }{\rho _{0}})^{G(x)}  \label{epotsym}
\end{equation}%
with $F(x=0)=129.981$ MeV, $G(x=0)=1.059$, $F(x=-1)=3.673$ MeV, and $%
G(x=-1)=1.569$. We note that the MID interaction reproduces very
well the EOS of isospin-asymmetric nuclear matter with the MDI
interaction at zero temperature for both $x=0$ and $x=-1$. The
single nucleon potential in the MID interaction can be directly
obtained as
\begin{equation}
U_{\text{MID}}(\rho ,\delta ,\tau )=\alpha \frac{\rho }{\rho _{0}}+\beta (%
\frac{\rho }{\rho _{0}})^{\gamma }+U^{\text{asy}}(\rho ,\delta ,\tau ),
\end{equation}%
with
\begin{eqnarray}
U^{\text{asy}}(\rho ,\delta ,\tau ) &=&\left[ 4F(x)\frac{\rho }{\rho _{0}}%
+4(18.6-F(x))(\frac{\rho }{\rho _{0}})^{G(x)}\right] {\tau }{\delta }  \notag
\\
&+&(18.6-F(x))(G(x)-1)(\frac{\rho }{\rho _{0}})^{G(x)}{\delta }^{2}.
\label{Uasy}
\end{eqnarray}%
Therefore, the single nucleon potential in the MID interaction is fully
momentum-independent. It also leads to the fact that the potential energy
density and the single nucleon potential in the MID interaction are
independent of the temperature.

The momentum-dependent part in the MDI interaction is also isospin dependent
while the MID interaction is fully momentum independent. In order to see the
effect of the momentum dependence of the isovector part of the single
nucleon potential (nuclear symmetry potential), we can introduce an
isoscalar momentum-dependent interaction, called extended MDYI (eMDYI)
interaction since it has the same functional form as the well-known MDYI
interaction for symmetric nuclear matter \cite{gale90}. In the eMDYI
interaction, the potential energy density $V_{\text{eMDYI}}(\rho ,T,\delta )$
of a thermally equilibrated asymmetric nuclear matter at total density $\rho
$, temperature $T$ and isospin asymmetry $\delta $ can be written as
\begin{eqnarray}
V_{\text{eMDYI}}(\rho ,T,\delta ) &=&\frac{A}{2}\frac{\rho ^{2}}{\rho _{0}}+%
\frac{B}{1+\sigma }\frac{\rho ^{1+\sigma }}{{\rho _{0}}^{\sigma }}  \notag \\
&+&\frac{C}{\rho _{0}}\int \int d^{3}pd^{3}p^{\prime }\frac{f_{0}(\vec{r},%
\vec{p})f_{0}(\vec{r},\vec{p}^{\prime })}{1+(\vec{p}-\vec{p}^{\prime
})^{2}/\Lambda ^{2}}  \notag \\
&+&{\rho }E_{sym}^{pot}(\rho ,x){\delta }^{2}.  \label{MDYIV}
\end{eqnarray}%
Here $f_{0}(\vec{r},\vec{p})$ is the phase space distribution function of
\emph{symmetric nuclear matter} at total density $\rho $ and temperature $T$%
. $E_{sym}^{pot}(\rho ,x)$ has the same expression as Eq.~(\ref{epotsym}).
We set $A=\frac{A_{u}+A_{l}}{2}$ and $C=\frac{C_{\tau ,-\tau }+C_{\tau ,\tau
}}{2}$, and $B$, $\sigma $ and $\Lambda $ have the same values as in the MDI
interaction, so that the eMDYI interaction gives the same EOS of asymmetric
nuclear matter as the MDI interaction at zero temperature for both $x=0$ and
$x=-1$. The single nucleon potential in the eMDYI interaction can be
obtained as%
\begin{equation}
U_{\text{eMDYI}}(\rho ,T,\delta ,\vec{p},\tau )=U^{0}(\rho ,T,\vec{p}%
)+U^{asy}(\rho ,\delta ,\tau ),
\end{equation}%
where
\begin{eqnarray}
U^{0}(\rho ,T,\vec{p}) &=&A\frac{\rho }{\rho _{0}}+B(\frac{\rho }{\rho _{0}}%
)^{\sigma }  \notag \\
&+&\frac{2C}{\rho _{0}}\int d^{3}p^{\prime }\frac{f_{0}(\vec{r},\vec{p})}{1+(%
\vec{p}-\vec{p}^{\prime })^{2}/\Lambda ^{2}}  \label{Usym}
\end{eqnarray}%
and $U^{\text{asy}}(\rho ,\delta ,\tau )$ is the same as Eq.~(\ref{Uasy})
which implies that the symmetry potential is identical for the eMDYI and MID
interactions. Therefore, in the eMDYI interaction, the isoscalar part of the
single nucleon potential is momentum dependent but the nuclear symmetry
potential is not. For symmetric nuclear matter, the single nucleon potential
in the eMDYI interaction is exactly the same as that in the MDI interaction.
We note that the same strategy has been used to study the momentum
dependence effects in heavy-ion collisions in a previous work \cite{chen04}.

At zero temperature, $f_{\tau }(\vec{r},\vec{p})$ $=\frac{2}{h^{3}}\Theta
(p_{f}(\tau )-p)$ and all the integrals in above expressions can be
calculated analytically, while at a finite temperature $T$, the phase space
distribution function becomes the Fermi distribution
\begin{equation}
f_{\tau }(\vec{r},\vec{p})=\frac{2}{h^{3}}\frac{1}{\exp (\frac{\frac{p^{2}}{%
2m_{_{\tau }}}+U_{\tau }-\mu _{\tau }}{T})+1}  \label{f}
\end{equation}%
where $\mu _{\tau }$ is the proton or neutron chemical potential and can be
determined from%
\begin{equation}
\rho _{\tau }=\int f_{\tau }(\vec{r},\vec{p})d^{3}p.
\end{equation}%
In the above, $m_{_{\tau }}$ is the proton or neutron mass and $U_{\tau }$
is the proton or neutron single nucleon potential in different interactions.
From a self-consistency iteration scheme \cite{gale90,xu07}, the chemical
potential $\mu _{\tau }$ and the distribution function $f_{\tau }(\vec{r},%
\vec{p})$ can be determined numerically.

From the chemical potential $\mu _{\tau }$ and the distribution function $%
f_{\tau }(\vec{r},\vec{p})$, the energy per nucleon $E(\rho ,T,\delta )$ can
be obtained as
\begin{equation}
E(\rho ,T,\delta )=\frac{1}{\rho }\left[ V(\rho ,T,\delta )+{\sum_{\tau }}%
\int d^{3}p\frac{p^{2}}{2m_{\tau }}f_{\tau }(\vec{r},\vec{p})\right] .
\label{E}
\end{equation}%
Furthermore, we can obtain the entropy per nucleon $S_{\tau }(\rho ,T,\delta
)$ as
\begin{equation}
S_{\tau }(\rho ,T,\delta )=-\frac{8\pi }{{\rho }h^{3}}\int_{0}^{\infty
}p^{2}[n_{\tau }\ln n_{\tau }+(1-n_{\tau })\ln (1-n_{\tau })]dp  \label{S}
\end{equation}%
with the occupation probability%
\begin{equation}
n_{\tau }=\frac{1}{\exp (\frac{\frac{p^{2}}{2m_{_{\tau }}}+U_{\tau }-\mu
_{\tau }}{T})+1}.
\end{equation}%
Finally, the pressure $P(\rho ,T,\delta )$ can be calculated from the
thermodynamic relation
\begin{eqnarray}
P(\rho ,T,\delta ) &=&\left[ T{\sum_{\tau }}S_{\tau }(\rho ,T,\delta
)-E(\rho ,T,\delta )\right] \rho  \notag \\
&&+\sum_{\tau }\mu _{\tau }\rho _{\tau }.  \label{P}
\end{eqnarray}

With the above theoretical models, we can now study the LG phase transition
in hot asymmetric nuclear matter. The phase coexistence is governed by the
Gibbs conditions and for the asymmetric nuclear matter two-phase coexistence
equations are
\begin{eqnarray}
\mu _{i}^{L}(T,\rho _{i}^{L}) &=&\mu _{i}^{G}(T,\rho _{i}^{G}),(i=n\text{
and }p)  \label{coexistencemu} \\
P_{i}^{L}(T,\rho _{i}^{L}) &=&P_{i}^{G}(T,\rho _{i}^{G}),(i=n\text{ or }p)
\label{coexistenceP}
\end{eqnarray}%
where $L$ and $G$ stand for liquid phase and gas phase, respectively. The
chemical stability condition is given by
\begin{equation}
\left( \frac{\partial {\mu }_{n}}{\partial {\delta }}\right) _{P,T}>0\text{
and }\left( \frac{\partial {\mu }_{p}}{\partial {\delta }}\right) _{P,T}<0.
\label{Cstability}
\end{equation}%
The Gibbs conditions (\ref{coexistencemu}) and (\ref{coexistenceP}) for
phase equilibrium require equal pressures and chemical potentials for two
phases with different concentrations. For a fixed pressure, the two
solutions thus form the edges of a rectangle in the proton and neutron
chemical potential isobars as a function of isospin asymmetry $\delta $ and
can be found by means of the geometrical construction method \cite%
{muller95,su00}.
\begin{figure}[tbh]
\includegraphics[scale=0.9]{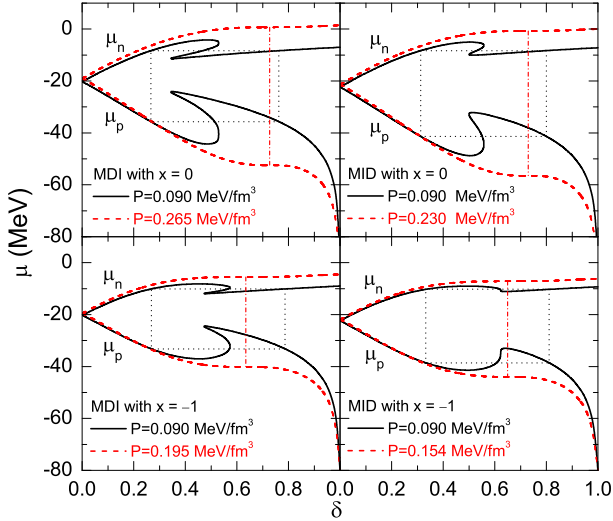}
\caption{{\protect\small (Color online) The chemical potential isobar as a
function of the isospin asymmetry $\protect\delta $ at $T=10$ MeV in the MDI
and MID interactions with $x=0$ and $x=-1$. The geometrical construction
used to obtain the isospin asymmetries and chemical potentials in the two
coexisting phases is also shown. }}
\label{mudeltamdimid}
\end{figure}

The solid curves shown in Fig.~\ref{mudeltamdimid} are the proton and
neutron chemical potential isobars as a function of the isospin asymmetry $%
\delta $ at a fixed temperature $T=10$ MeV and pressure $P=0.090$ MeV$/$fm$%
^{3}$ by using the MDI and MID interactions with $x=0$ and $x=-1$. The
resulting rectangles from the geometrical construction are also shown by
dotted lines in Fig.~\ref{mudeltamdimid}. For each interaction, the two
different values of $\delta $ correspond to two different phases with
different densities and the lower density phase (with larger $\delta $
value) defines a gas phase while the higher density phase (with smaller $%
\delta $ value) defines a liquid phase. Collecting all such pairs of $\delta
(T,P)$ and $\delta ^{\prime }(T,P)$ thus forms the binodal surface. From
Fig.~\ref{mudeltamdimid}, one can see that different interactions give
different shapes for the chemical potential isobar. When the pressure
increases and approaches the critical pressure $P_{\text{C}}$, an inflection
point will appear for proton and neutron chemical potential isobars, i.e.,%
\begin{equation}
\left( \frac{\partial {\mu }}{\partial {\delta }}\right) _{P_{\text{C}%
},T}=\left( \frac{\partial ^{2}{\mu }}{\partial {\delta }^{2}}\right) _{P_{%
\text{C}},T}=0.
\end{equation}%
Above the critical pressure, the chemical potential of neutrons (protons)
increases (decreases) monotonically with $\delta $ and the chemical
instability disappears. In Fig.~\ref{mudeltamdimid}, we also show the
chemical potential isobar at the critical pressure by the dashed curves. At
the critical pressure, the rectangle is degenerated to a line vertical to
the $\delta $ axis as shown by dash-dotted lines in Fig.~\ref{mudeltamdimid}%
. The values of the critical pressure are $0.265$, $0.230$, $0.195$ and $%
0.154$ MeV$/$fm$^{3}$ for the MDI interaction with $x=0$, MID interaction
with $x=0$, MDI interaction with $x=-1$ and MID interaction with $x=-1$,
respectively. It is interesting to see that the different interactions give
different values of the critical pressure. Especially, the stiffer symmetry
energy ($x=-1$) gives a smaller critical pressure while the inclusion of the
momentum dependence in the interaction (MDI) gives a larger critical
pressure.
\begin{figure}[tbh]
\includegraphics[scale=0.85]{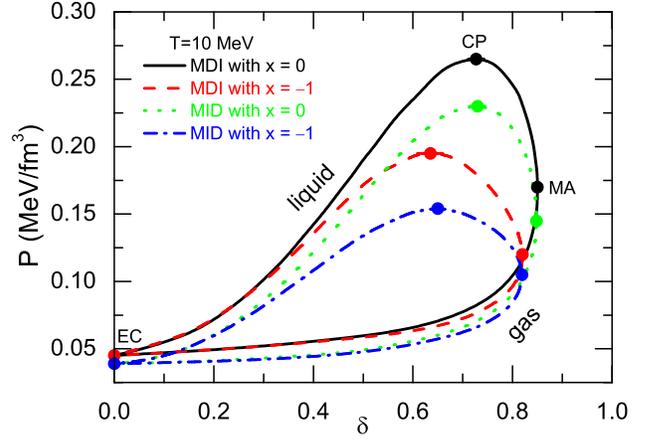}
\caption{{\protect\small (Color online) The section of binodal surface at $%
T=10$ MeV in the MDI and MID interactions with $x=0$ and $x=-1$. The
critical point (CP), the points of equal concentration (EC) and maximal
asymmetry (MA) are also indicated.}}
\label{Pdeltamdimid}
\end{figure}

In Fig.~\ref{Pdeltamdimid}, we show the section of the binodal surface at $%
T=10$ MeV for the MDI and MID interactions with $x=0$ and $x=-1$. On the
left side of the binodal surface there only exists a liquid phase and on the
right side only a gas phase exists. In the region of \textquotedblleft filet
mignon\textquotedblright\ is the coexistence phase of liquid phase and gas
phase. %Gas phase with a fixed isospin asymmetry can go through the
%coexistence region and reach its liquid phase when the nuclear matter is
%isothermally compressed.
Interestingly, we can see from Fig.~\ref%
{Pdeltamdimid} that the stiffer symmetry energy ($x=-1$) significantly
lowers the critical point (CP) and makes the maximal asymmetry (MA) a little
smaller. Meanwhile, the momentum dependence in the interaction (MDI) lifts
the CP in a larger amount, while it seems to have no effects on the MA
point. In addition, just as expected, the value of $x$ does not affect the
equal concentration (EC) point while the momentum dependence lifts it
slightly (by about $0.005$ MeV/fm$^{3}$). These features clearly indicate
that the critical pressure and the area of phase-coexistence region in hot
asymmetric nuclear matter is very sensitive to the stiffness of the symmetry
energy with a softer symmetry energy giving a higher critical pressure and a
larger area of phase-coexistence region. Meanwhile, the momentum dependence
in the interaction significantly increases the critical pressure and
enlarges the area of phase-coexistence region.
\begin{figure}[tbh]
\includegraphics[scale=0.9]{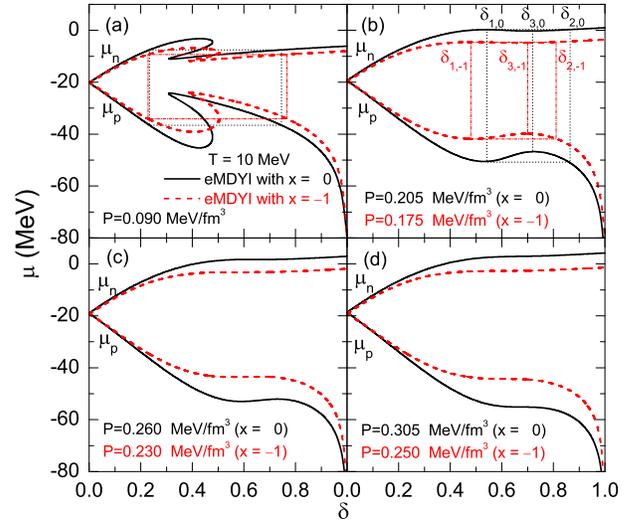}
\caption{{\protect\small (Color online) The chemical potential isobar as a
function of isospin asymmetry $\protect\delta $ at $T=10$ MeV in the eMDYI
interaction with $x=0$ and $x=-1$.}}
\label{mudeltamdyi}
\end{figure}
\newline

We now turn to the case of the eMDYI interaction. In the eMDYI interaction,
the resulting single nucleon potential is momentum dependent but its
momentum dependence is isospin-independent. Comparing the results with those
of the MDI interaction, we can extract information about the effects of the
momentum dependence of the symmetry potential while the effects of the
momentum dependence of the isoscalar part of the single nucleon potential
can be studied by comparing the results with those of the MID interaction.
Shown in Fig.~\ref{mudeltamdyi} is the chemical potential isobar as a
function of the isospin asymmetry $\delta $ at $T=10$ MeV by using the eMDYI
interaction with $x=0$ and $x=-1$. Compared with the results from the MDI
and MID interactions, the main difference is that the left (and right)
extrema of $\mu _{n}$ and $\mu _{p}$ do not correspond to a same $\delta $
but they do for the MDI and MID interactions as shown in Fig.~\ref%
{mudeltamdimid}. In particular, for the eMDYI interaction, the chemical
potentials of protons and neutrons are seen to exhibit asynchronous
variation with pressure, i.e., the chemical potential of neutrons increase
more rapidly with pressure than that of protons. This asynchronous variation
is uniquely determined by the special momentum dependence in the eMDYI
interaction within the present self-consistent thermal model. Actually, it
is this asynchronous variation that leads to the fact that the left (and
right) extrema of $\mu _{n}$ and $\mu _{p}$ correspond to different values
of $\delta $.

At lower pressures, for example, $P=0.090$ MeV/fm$^{3}$ as shown in Fig.~\ref%
{mudeltamdyi} (a), the rectangle can be accurately constructed and thus the
Gibbs conditions (\ref{coexistencemu}) and (\ref{coexistenceP}) have two
solutions. Due to the asynchronous variation of $\mu _{n}$ and $\mu _{p}$
with pressure, we will get a limiting pressure $P_{\lim }$ above which no
rectangle can be constructed and the coexistence equations (\ref%
{coexistencemu}) and (\ref{coexistenceP}) have no solution. Fig.~\ref%
{mudeltamdyi} (b) shows the case at the limiting pressure with $P_{\lim
}=0.205$ and $0.175$ MeV/fm$^{3}$ for $x=0$ and $x=-1$, respectively. In
this limit case, we note that for $x=0$ the left edge of the rectangle
actually corresponds to the left extremum of $\mu _{p}$ and the pair $\delta
_{1,0}=0.600$ and $\delta _{2,0}=0.750$ form the two edges of the rectangle
(the end of the binodal surface) while for $x=-1$ it corresponds to the left
extremum of $\mu _{n}$ and the pair $\delta _{1,-1}=0.480$ and $\delta
_{2,-1}=0.811$ form the two edges of the rectangle. The $\delta _{3,0}=0.720$
and $\delta _{3,-1}=0.700$ indicated in Fig.~\ref{mudeltamdyi} (b)
correspond to the maximum of $\mu _{p}$ for $x=0$ and $x=-1$, respectively.
With increasing pressure, namely, at $P=0.260$ and $0.230$ MeV/fm$^{3}$ for $%
x=0$ and $x=-1$, respectively, $\mu _{n}$ passes through an inflection point
while $\mu _{p}$ still has a chemically unstable region and this case is
shown in Fig.~\ref{mudeltamdyi} (c). When the pressure is further increased
to $P=0.305$ and $0.250$ MeV/fm$^{3}$ for $x=0$ and $x=-1$, respectively, as
shown in Fig.~\ref{mudeltamdyi} (d), $\mu _{p}$ passes through an inflection
point while $\mu _{n}$ increases monotonically with $\delta $. These
features indicate that the asynchronous variation of $\mu _{n}$ and $\mu
_{p} $ with pressure also depends on the value of $x$.
\begin{figure}[tbh]
\includegraphics[scale=0.85]{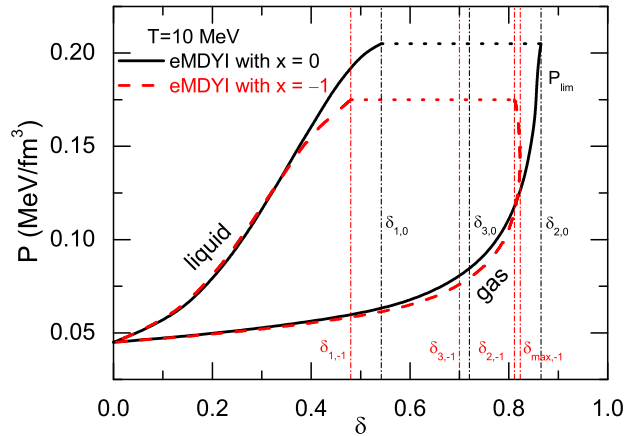}
\caption{{\protect\small (Color online) The section of binodal surface at $%
T=10$ MeV in the eMDYI interaction with $x=0$ and $x=-1$. $P_{\text{lim}}$
represents the limiting pressure and $\protect\delta $}$_{{\protect\small i,j%
}}${\protect\small 's are discussed in the text.}}
\label{Pdeltamdyi}
\end{figure}

Fig.~\ref{Pdeltamdyi} displays the section of the binodal surface at $T=10$
MeV by using the eMDYI interaction with $x=0$ and $x=-1$. We can see that
the curve is cut off at the limiting pressure with $P_{\lim }=0.205$ and $%
0.175$ MeV/fm$^{3}$ for $x=0$ and $x=-1$, respectively. It is interesting to
see that for $x=-1$ there is an bending point at $\delta _{max,-1}=0.823$,
but for $x=0$ there is no such point. For $x=-1$ the situation is similar to
that found in Ref.~\cite{su00} using different models and the isospin
asymmetry $\delta $ can be divided into four regions: $[0,\delta _{1,-1}]$, $%
[\delta _{1,-1},\delta _{3,-1}]$, $[\delta _{3,-1},\delta _{2,-1}]$ and $%
[\delta _{2,-1},\delta _{max,-1}]$. In the first region $[0,\delta _{1,-1}]$%
, the system begins at gas phase, experiences a LG phase transition and ends
at liquid phase. In the second region $[\delta _{1,-1},\delta _{3,-1}]$, it
begins at gas phase, enters a two-phase region and becomes unstable at the
limiting pressure, as chemical instability condition Eq.~(\ref{Cstability})
is destroyed in this region. In the third region $[\delta _{3,-1},\delta
_{2,-1}]$, the system will end in a stable phase as Eq.~(\ref{Cstability})
is satisfied in this region. In the fourth region $[\delta _{2,-1},\delta
_{max,-1}]$, the system enters and leaves the two-phase region on the same
branch, so it remains in the gas phase at the end. For $x=0$, the situation
is simpler and $\delta $ can be divided into three regions: $[0,\delta
_{1,0}]$, $[\delta _{1,0},\delta _{3,0}]$ and $[\delta _{3,-1},\delta
_{2,0}] $. The discussion is similar to that of the first, second and third
regions in the case of $x=-1$.

From Fig.~\ref{Pdeltamdyi}, we can also see that the limiting pressure and
the area of phase-coexistence region are still sensitive to the stiffness of
the symmetry energy with a softer symmetry energy ($x=0$) giving a higher
limit pressure and a larger area of phase-coexistence region. Comparing the
results of the MDI and MID interactions shown in Fig.~\ref{Pdeltamdimid}, we
can see that for pressures lower than the limiting pressure, the binodal
surface from the eMDYI interaction is similar to that from the MDI
interaction. This feature implies that the momentum dependence of the
symmetry potential has little influence on the LG phase transition in hot
asymmetric nuclear matter while the momentum dependence of the isoscalar
single nucleon potential significantly enlarges the area of
phase-coexistence region for pressures lower than the limiting pressure. For
pressures above the limiting pressure, the momentum dependence of both the
isoscalar and isovector single nucleon potentials becomes important.

In summary, we have studied the liquid-gas phase transition in hot
neutron-rich nuclear matter within a self-consistent thermal model
using three different nuclear effective interactions, namely, the
isospin and momentum dependent MDI interaction constrained by the
isospin diffusion data in heavy-ion collisions, the
momentum-independent MID interaction, and the isoscalar
momentum-dependent eMDYI interaction. At zero temperature, the
above three interactions give the same EOS for asymmetric nuclear
matter. The MDI interaction is realistic, while the MID and eMDYI
interactions are only used as references in order to explore
effects of the isospin and momentum dependence on the liquid-gas
phase transition. Since the symmetry energy corresponding to the
MDI interaction has been already restricted within a narrow range
by the isospin diffusion data in heavy-ion collisions, predictions
using the MDI interaction thus allow us to constrain the
liquid-gas phase boundary in asymmetric nuclear matter.

Comparing calculations with the three interactions, we find that
the boundary of the phase-coexistence region is very sensitive to
the density dependence of the nuclear symmetry energy. A softer
symmetry energy leads to a higher critical pressure and a larger
area of the phase-coexistence region. On the other hand, compared
with the momentum-independent MID interaction, the isospin and
momentum-dependent MDI interaction increases the critical pressure
and enlarge the area of phase-coexistence region. For the
isoscalar momentum-dependent eMDYI interaction, there is a
limiting pressure above which the liquid-gas phase transition
cannot take place due to the asynchronous variations of the
nucleon chemical potentials with pressure. Comparing the results
of the eMDYI interaction with those of the MDI and MID
interactions, we find that, for pressures lower than the limiting
pressure, the momentum dependence of the symmetry potential has
little influence on the liquid-gas phase transition in hot
asymmetric nuclear matter. While the momentum dependence of the
isoscalar single nucleon potential significantly enlarges the area
of phase-coexistence region. For pressures above the limiting
pressure, the momentum dependence of both the isoscalar and
isovector single nucleon potentials becomes important.

\begin{acknowledgments}
This work was supported in part by the National Natural Science Foundation
of China under Grant Nos. 10334020, 10575071, and 10675082, MOE of China
under project NCET-05-0392, Shanghai Rising-Star Program under Grant No.
06QA14024, the SRF for ROCS, SEM of China, the US National Science
Foundation under Grant No. PHY-0652548 and the Research Corporation.
\end{acknowledgments}

\end{document}